\documentclass[twocolumn]{revtex4-1}

\usepackage[T1]{fontenc}
\usepackage[latin9]{inputenc}
\usepackage{amsmath}
\usepackage{amssymb}
\usepackage{stmaryrd}
\usepackage{hyperref}  

\pdfpageheight\paperheight
\pdfpagewidth\paperwidth

\usepackage[english]{babel}
\title{On the physical implications of Hawking's spacetime foam}
\begin{document}
\author{Benjamin Schulz}
\email{Benjamin.Schulz@physik.uni-muenchen.de}
\affiliation{Department für Physik, Ludwig-Maximilians-Universität München, Theresienstraße 37, 80333 München}

\begin{abstract}
Hawking's spacetime foam model predicts that due to quantum fluctuations,
spacetime is filled with black hole like objects. We argue that Hawking's
model implies a cosmological constant of the observed order and that
it can also be used to solve the problem of time in quantum gravity.
\end{abstract}

\maketitle
\section{Introduction}

As is well known, the Hamiltonian of gravity is constrained and numerically
zero \cite{deWitt}. After canonical quantisation, this would lead
to observables whose expectation values are time independent. For
a system like this one may try to solve this so-called problem of
time with a canonical transformation which leads to a new non-zero
Hamiltonian and, after quantization, to a time dependent Wheeler DeWitt
(WDW) equation \cite{Peres}. However, Hajcek has shown in \cite{Haj}
that for the quantized FLRW universe, there exists no such canonical
transformation that yields a global time function. In the same article,
Hajcek also showed that choosing the wrong coordinate as a time gives
problems with unitarity. One may try to use this method locally. But
then one is faced with annother ambiguity. For simple systems, one
can show \cite{Kuchartime} that there are several different possible
canonical transformations that lead, in general, to different quantum
theories. 

In 1967, DeWitt argued in \cite{deWitt} that in a spacetime which
is asymptotically flat, one should add a non-vanishing boundary term
to the Hamiltonian constraint. DeWitt also writes that the addition
of non-zero boundary terms in the Wheeler DeWitt equation would result
in a dymamical theory and solve the problem of time. In 1983, Hartle
and Hawking showed \cite{Hartle} that the amplitude $\psi=\int\mathcal{D}g_{\mu\nu}e^{iS_{0}}$,
with the gravitational action $S_{\text{0}}=\frac{1}{16\pi}\int d^{4}x\sqrt{-g}R$
is an approximate solution of the WDW equation. Recently, Feng and
Matzner have made this result more precise for transition amplitudes
and they showed in \cite{Feng} that the problem of time is also there
in the amplitude of quantum gravity. If one does not add boundary
terms to the action, one is either left with a static system or one
does seem to have multiple choices for the time parameter which may
result in different quantum theories.

Another problem in cosmology is the cosmological constant problem.
Evaluating the vacuum vacuum amplitude $Z=\int\mathcal{D}\phi e^{iS(\phi)}$
for a matter field $\phi$ of mass m in a curved spacetime, one would
get an effective cosmological constant in the form of 
\begin{equation}
\Lambda_{eff}=\Lambda_{g}+\Lambda_{m}
\end{equation}
where $\Lambda_{g}$ is the bare cosmological constant from gravity
and $\Lambda_{m}$ is a matter contribution. The latter can be computed
from the amplitude as
\begin{equation}
\Lambda_{m}=+n_{i}\frac{m^{4}}{8\pi}\left(\frac{2}{\epsilon}-\gamma-ln\left(\frac{m^{2}}{4\pi\mu^{2}}\right)\right),
\end{equation}
where $n_{i}=\pm1$ for each bosonic/fermionic degree of freedom,
$\epsilon$ is a cut-off, $\gamma$ is the Euler-Mascheroni constant
and $\mu$ is a renormalization scale that is often set to $\mu=H_{0}$,
where $H_{0}$ is Hubble's constant \cite{everything,Cutoff,Shapiro}.
A suitable renormalization scheme can remove the cut off $\epsilon$
and one ends with 
\begin{equation}
\Lambda_{m}=n_{i}\frac{m^{4}}{8\pi}ln\left(\frac{m^{2}}{\mu^{2}}\right).
\end{equation}
We note that after the removal of the cut-off $\epsilon$, the matter
contribution to the cosmological constant crucially depends on the
value of $\mu$ such that different choices even can change the sign
of $\Lambda_{m}$ \cite{everything,Cutoff}. 

Additionally to the contributions from quantum fluctuations, $\Lambda_{m}$
would recieve contributions from electroweak and quantum chromodynamic
phase transitions if the entire standard model is taken into account.
The choice of $\mu=H_{0}$ leads to a matter contribution of around
$\Lambda_{m}\approx-9.7\cdot10^{8}GeV$ but the observed value of
$\Lambda_{eff}$ is of the order $\Lambda_{eff}\approx10^{-47}GeV$
\cite{Cutoff}. Currently, there is no universally accepted mechanism
that leads to an almost but not exactly cancellation of $\Lambda_{m}$,
although there are various proposals. This is called the cosmological
constant problem. From cosmological measurements, one observes that
\begin{equation}
\Lambda_{eff}=3\Omega_{\Lambda}H_{0}^{2}\approx2.13H_{0}^{2}
\end{equation}
 where $\Omega_{\Lambda}=0.71$ is the density parameter of the vacuum.
The problem why the cosmological constant is approximately equal to
the Hubble radius is the so called coincidence problem.

There are numerous attempts to solve this cosmological constant problem.
For example, there are the classic articles from Coleman \cite{Coleman},
Klebanov, Suesskind and Banks \cite{Klebanov} or Preskill \cite{Preskill}
that attempt to solve the cosmological constant problem with wormholes
but they do not get its numerically correct value. More recently,
there are articles from Carlip \cite{Spacetimefoamcarlip}, Padmanabhan
\cite{Padmanabahnfoam}, Xue \cite{Xue} or the proposal \cite{Cyriac}
that all follow a similar path. These papers, however, do not solve
the problem of time or the coincidence problem, and they are often
using additional assumptions that do not entirely come from the quantum
gravity amplitude. 

Then, there are works like \cite{Spacetimefoamccgerat} or the recent
proposal \cite{Cosmictime}, which suggest the cosmological constant
could have something to do with boundary terms of instantons and the
problem of time, but these papers fails to provide a precise value
of the constant. In another attempt, Barrow and Shaw \cite{constant}
threat the cosmological constant as a variable field in quantum gravity. They get its value, but only in the case of a universe with boundary terms and they do not provide a detailed mechanism why the cosmological constant should be threated in this way.

It is the purpose of this article to show that Hawking's spacetime
foam model from \cite{foam} can be easily adapted such that it solves
the problem of time in quantum gravity, the cosmological constant
problem and the coincidence problem simultaneously. The model remains
close to the amplitude of quantum gravity without needing much additional assumptions apart from quantum field theory in curved spacetime. 

Hawking's model leads to a spacetime filled with a gas of black holes.
As noted by Hawking in \cite{Bubble1,Bubble2,Bubble3} this could
have severe experimental consequences as it may led to the trajectories of particles which differ from the observed ones. In the last section, we therefore try to outline a proposal that might cure this.

\section{Hawking's model on spacetime foam revisited}

In his work on spacetime foam \cite{foam}, Hawking considers the
Euclidean action $I$ of gravity with cosmological constant 
\begin{eqnarray}
I & = & \frac{-1}{16\pi}\int d^{4}x\sqrt{g}(R-2\Lambda)\nonumber \\
 & = & -\frac{V\Lambda}{8\pi}.\label{eq:action2}
\end{eqnarray}
Using dimensional arguments, one can write
\begin{equation}
\sqrt{V}=-f/\Lambda,
\end{equation}
where $f$ is some dimensionless factor which can be positive or negative.
The action then becomes $I=-\frac{f^{2}}{8\pi\Lambda}$. In his article,
Hawking argues that $f^{2}=d^{2}\chi$ with $d$ as a numerical constant
and $\chi$ as the Euler number. This claim is based on the following
arguments: Hawking noted that for a spacetime which fulfills Einstein's
field equations in vacuum with cosmological constant, the Euler number
can be computed by 
\begin{equation}
\chi=\frac{1}{32\pi^{2}}\int d^{4}x\sqrt{g}(C_{\alpha\beta\gamma\mu}C^{\alpha\beta\gamma\mu}+\frac{8}{3}\Lambda^{2}),\label{eq:euler1}
\end{equation}
where $C_{\alpha\beta\gamma\mu}$ is the Weyl tensor. For large Euler
number, one possibility is a large $f^{2}=\Lambda^{2}V$. Hawking
claims that classically, one must have $f\geq-\sqrt{24}\pi$ and that
$f$ reaches its lower bound for an $S^{4}$ spacetime. This implies
for large $f^{2}=\Lambda^{2}V$ that $\Lambda$ must be negative.
On the other hand, if $\int d^{4}x\sqrt{g}C_{\alpha\beta\gamma\mu}C^{\alpha\beta\gamma\mu}$
is large, then this has a converging effect on the geodesics. Hawking
argues that there should not be conjugate points between any two points
in space. In order to prevent the geodesics to converge, he puts in
a negative $\Lambda$ such that the both terms in eq. (\ref{eq:euler1})
are of the same order, and one can write $f^{2}=\Lambda^{2}V\approx d^{2}\chi$.

For our universe, this argument is problematic since it only holds
for large negative $\Lambda$ whereas the observed cosmological constant
is small and positive. Furthermore, as Hawking writes in \cite{foam}
that his spacetime foam model describes a gas of black holes. Behind
the event horizon of black holes, one should find a large curvature
with converging geodesics. Additionally, Hawking later gets a stationary
point for $\Lambda$ which is positive, but he still continues to
use formulas based on $f^{2}=d^{2}\chi$. 

In order to avoid this problem, we will not write the action in the
form $I=-\frac{d^{2}\chi}{8\pi\Lambda}$ that Hawking uses, but we
will repeat some of his calculations with the usual Euclidean action
$I=-\frac{\Lambda V}{8\pi}$. Then we will add matter terms to the
model and look at their effect on the cosmological constant. Finally,
we will discuss the Hamiltonian that arises from this theory.

Hawking notes that one can write the amplitude of quantum gravity
as\begin{equation}
Z(\Lambda)=\int\mathcal{D}g_{\mu\nu}e^{-I}=\sum_{n}\langle g_{n}|e^{-\frac{\Lambda V}{8\pi}}|g_{n}\rangle\label{eq:wefewcwcewvccw}
\end{equation}
where the sum goes over a complete orthonormal basis $|g_{n}\rangle$
of the gravitational field. The amplitude in Euclidean quantum gravity
corresponds to a canonical partition sum. Hawking argues that this
partition sum is equivalent to
\begin{equation}
Z(\Lambda)=\int_{0}^{\infty}N(V)e^{-\frac{\Lambda V}{8\pi}}dV.\label{eq:Volumesummation}
\end{equation}
The function $N(V)$ describes the number of states between $V$ and
$V+dV$ and can be computed by the inverse Laplace transform of the
partition function $Z(\Lambda)$. Evaluating the complex integral
of the inverse Laplace transform with a saddle point evaluation 
as one does it in ordinary thermodynamics \cite{Laplacetrafo} leads
to
\begin{equation}N(V) \approx Z(\Lambda_s)e^{\frac{\Lambda_s V}{8\pi}}.\label{eq:fdshhlkorewdewqdx}
\end{equation}where the saddle point $\Lambda_s$ is given by $\left.\frac{d}{d\Lambda}\left(ln(Z)+\frac{\Lambda_s V}{8\pi}\right)\right|_{\Lambda_s}=0.$ The entropy of a system with $N$ microstates is then given by 
$S=k_{b}ln(N)\label{eq:entropy}$ where $k_{b}$ is Boltzmann's constant. 

The WKB method can be used to compute the amplitude up to the one
loop approximation and yields $Z=Ce^{-I}$. The one loop amplitude
of quantum gravity can also be calculated by Zeta function renormalisation
techniques \cite{Indefiniteness}. There, one expresses the perturbative
quantum corrections as determinants of operators. These determinats
diverge since the eigenvalues $\lambda_{n}$ of these operators increase
without bounds. Nevertheless, one can define these determinants to
be equal to $e^{-\frac{d\zeta}{ds}|_{s=0}}$ where $\zeta(s)=\sum_{n}\lambda_{n}^{-s}$
is the generalized zeta function, which is finite in analytic continuation
at $s=0$. Using $\Lambda^{2}V=f^{2}$, the cosmological constant
can be seen as a factor that rescales the action from eq. (\ref{eq:action2})
according to 
\begin{equation}
I=-\frac{\Lambda V}{8\pi}=\frac{1}{\Lambda}I_{0},
\end{equation}
where $I_{\text{0}}=-\frac{f^{2}}{8\pi}$. The works of Gibbons Hawking
and Perry \cite{Indefiniteness} show that if the action is rescaled
by $1/\Lambda$ then, the eigenvalues of the determinants in the Zeta
function renormalization scheme are multiplied by $\Lambda$. The
amplitude then scales with $C\propto\left(\frac{\Lambda}{\mu^{2}}\right)^{-\frac{1}{2}\gamma}$
where $\gamma$ is given in \cite{PerryScaling} for spacetimes with
cosmological constant as
\begin{equation}
\gamma=\frac{1}{16\pi^{2}}\left(\frac{848}{720}\int d^{4}x\sqrt{g}R_{\alpha\beta\gamma\delta}R^{\alpha\beta\gamma\delta}+\frac{438}{45}\Lambda^{2}V\right).
\end{equation}
Using the generalized Gauss Bonnet theorem, the Euler characteristic
for a compact manifold of dimension 4 can be computed by 
\begin{equation}
\chi=\frac{1}{32\pi^{2}}\int d^{4}x\sqrt{g}\left(R_{\alpha\beta\gamma\delta}R^{\alpha\beta\gamma\delta}-4R_{\mu\nu}R^{\mu\nu}+R^{2}\right).
\end{equation}
If Einstein's field equations in vacuum hold, then $R_{\mu\nu}=\Lambda g_{\mu\nu}$
and $R=4\Lambda$. The Euler number becomes 
\begin{equation}
\chi=\frac{1}{32\pi^{2}}\int d^{4}x\sqrt{g}R_{\alpha\beta\gamma\delta}R^{\alpha\beta\gamma\delta},
\end{equation}
and the factor $\gamma$ is then
\begin{equation}
\gamma=\frac{106}{45}\chi+\frac{73}{120\pi^{2}}\Lambda^{2}V.
\end{equation}
Finnally, the amplutude $Z=Ce^{-I}$ can be written as
\begin{equation}
Z(\Lambda,\chi)=\left(\frac{\Lambda}{\mu^{2}}\right)^{-\frac{1}{2}\left(\frac{106}{45}\chi+\frac{73}{120\pi^{2}}\Lambda^{2}V\right)}e^{\frac{\Lambda V}{8\pi}}.
\end{equation}
Relations for the cosmological constant can then be computed from the saddle point $\Lambda_s$ as indicated above and by maximizing 
the entropy for other parameters, like the Euler number $\chi$. Hawking's model of spacetime
foam is similar to the proposal of Barrow and Shaw \cite{constant},
who also consider $\Lambda$ to be a continuous field. A main difference
is that the calculations of Barrow and Shaw need a universe with an
apparent horizon that can be described with a GHY boundary term. This
is not necessary in Hawking's model.

In general, the dominant contributions to a path integral should come
from classical paths. If we use an amplitude $Z=Ce^{-I}$, one gets
from Equation (\ref{eq:fdshhlkorewdewqdx}) 
\begin{equation}
S=k_{b}\left(-I+\frac{\Lambda V}{8\pi}+\ln(C)\right).
\end{equation}
$\ln(C)$ comes from a perturbative expansion of the metric up to
one loop and can be seen as a quantum correction to the entropy that
one gets from the classical background. In order for the entropy not
to be affected much by this quantum correction, one should have $C\approx1$.
Since $C=\left(\frac{\Lambda}{\mu^2}\right)^{-\frac{106}{90}\chi-\frac{73}{240\pi^2}\Lambda^2V}$, $C\approx 1$ is consistent with
\begin{equation}
\Lambda\approx\mu^{2}.\label{eq:Lambdamu}
\end{equation}

We will consider the 4 volume as a function of $\Lambda$ by substituting $V(\Lambda)=\frac{f^2}{\Lambda^2}$ in the amplitude. If we then compute the saddle point $\Lambda_{s1}$ from \begin{equation}\left.\frac{\partial\left(\ln{Z(\Lambda)}+\frac{f^2}{8\pi\Lambda}\right)}{\partial\Lambda}\right|_{\Lambda_{s1}}=0\label{eq:SaddlepointLambda}\end{equation} we get the following expression \begin{equation}\Lambda_{s1}=-\frac{180\pi f^2}{848\pi^2\chi+219f^2}\label{eq:Lambdastationary1}\end{equation}

Substituting $f^2=\Lambda^2V$ back into eq. (\ref{eq:Lambdastationary1}) yields \begin{equation}\Lambda_{s1}=\frac{-90\pi}{219}\pm \frac{2\sqrt{2025\pi^2 V^2-46428\pi^2 V\chi}}{219V}\label{eq:Lambdas12}\end{equation}This formula for the stationary point is a bit different from the result of Hawking since we use a slightly different action. Furthermore, Hawking  also appears to make the substitution $\Lambda V=f^2/\Lambda$ only in the amplitude and not in the entire eq. (\ref{eq:SaddlepointLambda}) which leads to different signs for the expressions within in the square root.

The different signs in front of the square roots from eq. (\ref{eq:Lambdas12}) come from the $\Lambda^{2}$ contribution in the amplitude. They are a result of the usual $R^{2}$ corrections from quantum field theory in curved spacetimes \cite{DeWittcurved}. The formula with the negative sign before the square root has $\Lambda_{s1}=-\frac{60\pi}{73}$ in the limit $\chi=0$ and for $V\rightarrow\infty$. Since for our universe, the cosmological constant is positive, this solution appears to be unphysical. The formula with the positive sign has $\Lambda_{s1}=0$ in the limit $V\rightarrow\infty$ and for $\chi=0$. On the other hand, in the Limit $V=0$, the cosmological constant approaches $\Lambda_{s1}\rightarrow\infty$. If we set V to the 4 volume of the current observable universe, one sees that for a large negative Euler number, the current cosmological constant can be computed from the formula with the negative sign. 

If we would do the computation of the stationary point of $\Lambda$ with an amplitude where the $\Lambda^2 V$ term is omitted, we would get \begin{equation}\Lambda_{s0}=-\frac{212\pi\chi}{45V}.\end{equation} Plotting $\Lambda_{s1}$ and $\Lambda_{s0}$ shows that $\Lambda_{s1}<\Lambda_{s0}\forall V$ if we use the same $\chi$ as parameter. For the amplitude with the $\Lambda^2V$ term, the cosmological constant strives to zero much faster for large volumes.

We observe from eq. (\ref{eq:Lambdas12}) that $\Lambda_{s1}$ is not defined at all for $V<\frac{15476}{675}\chi$ if $\chi>0$. Solving eq. (\ref{eq:Lambdas12}) for $\chi$ yields \begin{equation}\chi=\frac{-3}{848\pi}\Lambda_{s1}(73\Lambda_{s1}\pi+60)V\equiv cV\label{eq:constantc}\end{equation} and we note that the $\chi>0$ holds only for $-\frac{60}{73\pi}<\Lambda_{s1}<0$. The Euler number is given by the alternating sum $\chi=\sum_{i=0}^4(-1)^i b_i$ where $b_0$ is the number of connected components and $b_1,b_2,b_3,b_4$ are the numbers of $i+1$ dimensional cavities. For a compact spacetime, $b_0=b_4=1$ and if the manifold is simply connected then $b_1=b_3$. A negative Euler number can therefore only occur for spacetimes which are not simply connected. This is the case for wormholes, for example. Since the Euler number is an integer eq. (\ref{eq:constantc}) implies that the volume of a compact spacetime where the gravitational entropy is at its maximum, is quantized. 

Despite some differences to Hawking's formulas, eq. (\ref{eq:constantc}) still describes a spacetime filled with one cavity per
$|c|^{-1}$ unit Planck volumes. However, the cavity density $|c|$ is now much smaller than the density we get from Hawking's article. In one Planck 4 volume, Hawking gets one gravitational cavity. With
eq. (\ref{eq:constantc}) and $\Lambda=5.6\cdot10^{-122}$ one gets
a number density of only $\approx3.79\cdot10^{-123}$ cavities per
unit Planck 4 volume. But on macroscopic scales, this is still a large
number density. If one considers a 3 Volume of a single cubic meter
for a time of one second (in si units), then, $\chi\approx1.62\cdot10^{25}$ cavities should have been produced during that second in this 3 volume.

With the stationary point from eq. (\ref{eq:Lambdastationary1}), one can compute the entropy \begin{equation}S=\ln{Z(\Lambda_{s1})}+\frac{f^2}{8\pi\Lambda_{s1}}.\label{eq:saddle}\end{equation}We can calculate the  Euler number that maximizes the entropy from \begin{multline}\frac{\partial S}{\partial\chi}=-\frac{106}{45}\left(\ln{(2)}+\ln{(3)}\right)-\frac{53}{45}\left(\ln{(5)}+\ln{(\pi)}\right)\\
-\frac{53}{45}\ln{\left(\frac{-f^2}{\left(848\pi^2\chi+219f^2\right)\mu^2}\right)}=0\end{multline}Solving this equation for $\mu$ implies $\mu^2=\Lambda_{s1}$ and one easily checks that \begin{equation}\frac{\partial^2 S}{\partial\chi^2}=\frac{44944}{45(219f^2+848\chi)}\end{equation} which means that this can be a maximum only for negative Euler numbers $\chi<\frac{-219f^2}{848\pi^2}$. 

If we go a different route and substitute $\chi$ in the amplitude by inserting eq. (\ref{eq:constantc}) and write the amplitude as a function of $\Lambda_{s1}$ only, we get \begin{equation}\frac{\partial S}{\partial \Lambda_{s1}}=-\frac{1}{4}\frac{f^2\ln{\left(\frac{\Lambda_{s1}}{\mu^2}\right)}}{\pi\Lambda_{s1}^2}\end{equation} which also shows that $\Lambda_{s1}=\mu^2$ is a maximum since \begin{equation}\frac{\partial^2 S}{\partial \Lambda_{s1}^2}=\frac{1}{4}\frac{f^2\left(2\ln{\left(\frac{\Lambda_{s1}}{\mu^2}\right)}-1\right)}{\pi\Lambda_{s1}^3}<0\end{equation} for a positive $\Lambda_{s1}=\mu^2$.

One should note that in Hawking's original article, eq. (\ref{eq:Lambdamu}) does not follow conclusively as the only possibility because of the different action that he is using. 

The coincidence problem of the value of the cosmological constant is now translated into the problem of a choice for the scale parameter $\mu$. In his original article, Hawking sets the renormalization scale $\mu$ to the Planck
length $l_{p}$ since he assumes that quantum gravity breaks down
at this scale. However, this choice is physically problematic. In
quantum field theory, renormalization scales are not cut offs that
should run to infinity or to the point where the theory becomes undefined.
Instead, these parameters should be determined with input from measurement.

According to the more recent literature on this topic, $\mu$ should be of the order of the energy of the gravitons involved in the amplitude\cite{Shapiro,Cutoff,everything}.
This energy can be estimated with an argument by Shapiro and Sola in \cite{Shapiro}. 

From Einstein's equation with matter and an energy momentum tensor $T_{\mu\nu}$,
one should expect $\mu\approx\sqrt{T_{\mu}^{\mu}}$
and it follows from Einstein's equation that $T_{\mu}^{\mu}\approx R$ if we take the input from measurement that
$\Lambda$ is very small. 

Currently, we live in a Friedmann cosmos with Hubble constant
\begin{equation}
H_{0}^{2}=\left(\frac{\dot{a}}{a}\right)^{2}=\frac{8\pi}{3}\left(\rho+\Lambda-\frac{k}{a^{2}}\right),\label{eq:friedmann}
\end{equation}
where $a$ is a time dependent scale factor, $\rho=\rho_{m}+\rho_{R}$ is the energy density of matter and
radiation. We set $k=0$ because
the universe is very flat at present. From experiment,
one knows that $\Lambda$ and $\rho$ are of the same order as the
critical density $\rho_{c}=\frac{3 H_0^2}{8\pi}$, therefore $H_{0}\approx\sqrt{\rho_{c}}$,
but one also has $T_{\mu}^{\mu}\approx\rho_{c}$, or
\begin{equation}
\mu\approx\sqrt{T_{\mu}^{\mu}}\approx\sqrt{R}\approx H_{0}\label{eq:scaleparameter}
\end{equation}

Using eq. (\ref{eq:scaleparameter}) in eq. (\ref{eq:Lambdamu}) that we computed from the amplitude implies
\begin{equation}
\Lambda=H_{0}^{2},\label{eq:fjkurztjwwr}
\end{equation}
which is approximately what is observed. 

The cosmological constant that one gets from the amplitude of quantum gravity is therefore perfectly compatible with the relation $H_0^2\approx\Lambda$ that emerges from the classical eq. (\ref{eq:friedmann}) for a universe where $k$ and $\rho$ can be neglected. This may be a step to solve the coincidence problem in quantum gravity. However, according to quantum field theory, the value of the scale parameter must ultimately be determined with input from measurement. Showing that $\Lambda=\mu^2=H_0^2$ just makes it necessary to measure $H_0$.

\section{Matter corrections to $\Lambda$}

In the following, we will describe what happens when we add matter
corrections. It is known that in curved spacetimes, after the application
of a suitable renormalization method, the effective Euclidean matter
action for a field with mass $m$ becomes 
\begin{equation}
I_{m;eff}=-n_{i}A\int d^{4}x\sqrt{g}\left(-\frac{1}{2}m^{4}a_{0}+m^{2}a_{1}-a_{2}\right),
\end{equation}
where $n=\pm1$ for each bosonic/fermionic degree of freedom, $a_{i}$
are the Seley DeWitt coefficients that depend on the matter fields
and on the topology and $A=\frac{1}{32\pi{}^{2}}\ln\left(\frac{m^{2}}{\mu^{2}}\right)$
has been computed with some renormalization method of choice. We can
then use this effective matter action to write a combined matter gravity
amplitude 
\begin{equation}
\tilde{Z}=Ce^{-\tilde{I}}\label{eq:Amplitude}
\end{equation}
 with $\tilde{I}=I_{gravity}+I_{m,eff}$.

In general, $a_{0}=1$ and $a_{1}\propto R$. Therefore, the first
term in $I_{m,eff}$ can be written as $\frac{1}{8\pi}\int d^{4}x\sqrt{g}\Lambda_{m}$
with 
\begin{equation}
\Lambda_{m}=n\frac{8\pi}{2}m^{4}a_{0}A=n\frac{m^{4}}{8\pi}\ln\left(\frac{m^{2}}{\mu^{2}}\right).
\end{equation}
It is a topology independent renormalization of the cosmological constant,
which then becomes a sum
\begin{equation}
\Lambda_{eff}=\Lambda_{g}+\Lambda_{m},
\end{equation}
where $\Lambda_{g}$ is the cosmological constant that we get from
gravity alone. If we set $a_{2}=0$, then the effective action will
only correspond to an additional cosmological constant term and a
renormalization of Newton's constant. The scaling behavior of the
gravity amplitude was derived in \cite{PerryScaling} for solutions
with an arbitrary cosmological constant. These results still hold
if we set $a_{2}=0$ in the matter amplitude, but one has to replace
$\Lambda$ by $\Lambda_{eff}$ everywhere. Before we compute the stationary point, we have to substitute $\Lambda_{eff} V=f^2/\Lambda_{eff}$ in the amplitude. Since the contribution
$\Lambda_{m}$ is not a variable field, we can compute the new stationary
value $\Lambda_{eff,s0}$ of the cosmological constant from solving \begin{equation}\frac{d}{d\Lambda_{g}}\left(\frac{f^2}{8\pi\Lambda_{eff}}+\ln{(\tilde{Z}(\Lambda_{eff}))}\right)=0\label{eq:stationary2}\end{equation} for $\Lambda_{eff}$
and the Euler number at maximum entropy from solving $\frac{d}{d\chi}\tilde{S}(\Lambda_{eff,s0})=0$ for $\chi$ where $\tilde{S}(\Lambda_{eff,s0})\approx\ln{(\tilde{Z}(\Lambda_{eff,s0}))}+\frac{f^2}{8\pi\Lambda_{eff,s0}}$. Doing this computation shows that the relations between  $\Lambda_{eff,s0}$ and the Euler characteristic and the volume are again given by eq. (\ref{eq:Lambdastationary1}). Similarly, we also get the value $\Lambda_{eff,s0}=\mu^{2}$ from solving $\frac{d}{d\chi}\tilde{S}(\Lambda_{eff,s0})=0$ for $\mu$. Therefore, the contribution of $\Lambda_{m}=-9.7\cdot10^{8}GeV$ from the standard model matter has no observable effect at all. 

This becomes different if we include the contributions of terms $\propto-\int d^{4}x\sqrt{g}a_{2}$
in the matter action. For example, for a scalar boson field, one has

\begin{multline}
a_{2Boson}=\frac{1}{180}R_{\mu\nu\alpha\beta}R^{\mu\nu\alpha\beta}-\frac{1}{180}R^{\mu\nu}R_{\mu\nu}\\
+\frac{1}{2}\left(\frac{1}{6}-\zeta\right)^{2}R^{2}-\frac{1}{6}\left(\frac{1}{5}-\zeta\right)\Box R,
\end{multline}
where $\zeta$ is an undetermined coupling parameter that has to be
measured by experiment. Similarly, for fermions, if we assume they
have 4 spinor components and take the trace of them, one gets
\begin{multline}
a_{2Fermion}=-\frac{7}{360}R_{\mu\nu\alpha\beta}R^{\mu\nu\alpha\beta}-\frac{1}{45}R^{\mu\nu}R_{\mu\nu}\\
+\frac{1}{72}R^{2}+\frac{1}{30}\boxempty R.
\end{multline}
Obviously the effective matter action is proportional to the Euler
number:

\begin{equation}
I_{m,eff}\propto-\int d^{4}x\sqrt{g}a_{2}\propto-\eta\chi,
\end{equation}
where $\eta$ is some factor that depends on the particle masses and
which is positive for bosons and negative for fermions. 

The energy momentum tensor $T_{\mu\nu}$ can be computed from the
variation of the matter action. Its trace $T$ gets into Einstein's
field equations and leads to $R=4\Lambda-8\pi T$ . For a conformal
massless model, it is known that one gets $T=-\frac{a_{2}}{16\pi}$.
In that case, one therefore has an additional term $\propto\chi$
in the gravity action
\begin{equation}
I=\frac{-\Lambda V}{8\pi}-\frac{1}{2}\int d^{4}x\sqrt{g}a_{2}.
\end{equation}

If the particles have mass, the energy momentum tensor becomes more
complicated than in the massless case. The variation of the effective
matter action with a nonvanishing $a_{2}$ term is given by: 

\begin{align}
\frac{\langle out,0|T_{\mu\nu}|0,in\rangle}{\langle out,0|0,in\rangle} & =a^{\;\;(1)}H_{\mu\nu}+b^{\;\;(2)}H_{\mu\nu}+c^{\;\;(3)}H_{\mu\nu}\nonumber \\
 & =\tilde{a}^{\;\;(1)}H_{\mu\nu}+\tilde{b}^{\;\;(2)}H_{\mu\nu}\label{eq:modifiedeinstein}
\end{align}
In eq. (\ref{eq:modifiedeinstein}), $a,b,c$ are numerical coefficients
that must be determined by measurement and one has $^{(1)}H_{\mu\nu}=\frac{1}{\sqrt{g}}\frac{\delta}{\delta g^{\mu\nu}}\int d^{4}x\sqrt{g}R^{2}$,
$^{(2)}H_{\mu\nu}=\frac{1}{\sqrt{g}}\frac{\delta}{\delta g^{\mu\nu}}\int d^{4}x\sqrt{g}R^{\alpha\beta}R_{\alpha\beta}$
and $^{(3)}H_{\mu\nu}=\frac{1}{\sqrt{g}}\frac{\delta}{\delta g^{\mu\nu}}\int d^{4}x\sqrt{g}R^{\alpha\beta\gamma\delta}R_{\alpha\beta\gamma\delta}$.
One of the three tensors in eq. (\ref{eq:modifiedeinstein}) can be
absorbed into the other two with new coefficients $\tilde{a},\tilde{b}$
since the variation of the Euler number vanishes because it is a topological
invariant. After computing the variation, the tensors do not seem
to be a simple function of the Euler number. Therefore, they do not
seem to affect derivatives $\frac{dS}{d\chi}$ from which we computed
the value of the cosmological constant. Nevertheless, they could change
the scaling behavior of the amplitude after quantisation. In order
to determine that, one would have to quantize $T$ with zeta functions.
Unfortunately, it is known that severe problems arise if one tries
to quantize $R^{2}$ modifications of gravity \cite{Globalap}. One
can make the argument that $\tilde{a},\tilde{b}$ must be very small
as we do not observe corrections from $^{(1)}H_{\mu\nu}$ and $^{(2)}H_{\mu\nu}$
of Einstein's equations in the macroscopic world. This line of reasoning
is problematic in quantum gravity, since the latter also deals with
very small corrections to classical physics. If one uses this argument
nevertheless, then, the trace $T$ in the gravity action, and the
use of $\langle out,0|T_{\mu\nu}|0,in\rangle$ in Einstein's field
equations can be neglected. The resulting equations of motion yield
a spacetime that fulfills Einstein's equations in vacuum with cosmological
constant. With $R=4\Lambda$ and $R_{\mu\nu}=\Lambda g_{\mu\nu}$,
one computes
\begin{equation}
\int d^{4}x\sqrt{g}a_{2}=\frac{45}{8\pi^{2}}\chi-\frac{1}{45}\Lambda^{2}V+\left(\frac{1}{6}-\zeta\right)^{2}8\Lambda^{2}V
\end{equation}
 for bosons and for fermions, one arrives at
\begin{equation}
\int d^{4}x\sqrt{g}a_{2}=-\frac{45}{28\pi^{2}}\chi-\frac{4}{45}\Lambda^{2}V+\frac{2}{9}\Lambda^{2}V,
\end{equation}
 where we used that the integral over $\square R$ vanishes.

Hence, if we ignore potential effects of the $R^{2}$ terms for the
scaling behavior of the gravity amplitude, we can expect that the
combined matter gravity action yields an amplitude of the form 
\begin{equation}
\overline{Z}(\Lambda_{eff},\chi)=\left(\frac{\Lambda_{eff}}{\mu^{2}}\right)^{-\frac{53}{45}\chi+\frac{73}{240\pi^{2}}\Lambda_{eff}^{2}V}e^{\frac{V\Lambda_{eff}}{8\pi}+\eta\chi+\gamma\Lambda_{eff}^{2}V}
\end{equation}
 where $\eta$ and $\gamma$ are some numerical factors that depend
on particle masses. If we are substituting $ V(\Lambda_{eff})=f^2/\Lambda_{eff}^2$ in this amplitude and compute the saddle point $\Lambda_{eff,s1}$ with the same method as in eq. (\ref{eq:stationary2}), we get the same relations between $\Lambda_{eff,s1}$ and the volume and Euler number as we computed them in eq. (\ref{eq:Lambdastationary1}) from the amplitude without the $e^{\gamma\Lambda^2V}$ and $e^{\eta\chi}$ terms. 

With $\Lambda_{eff,s1}$ given by eq. (\ref{eq:Lambdastationary1}),  we can write the entropy as $\overline{S}(\Lambda_{eff,s1})\approx\ln{(\overline{Z}(\Lambda_{eff,s1}))}+{\frac{f^2}{8\pi\Lambda_{eff,s1}}}$. If we solve $\frac{d\overline{S}}{d\chi}=0$ for $\mu^2$ we get
\begin{equation}
\Lambda_{eff,s1}=e^{\frac{45}{53}\eta}\mu^{2}.
\end{equation}and one checks that this is a maximum for large negative Euler numbers as before.

The factor $\eta$ has different signs for fermions and bosons and
depends on their masses. Thereby, with the addition of appropriate
matter particles, one may e.g. correct the stationary value such that
it becomes $\Lambda=2.13H_{0}^{2}$ if we set $\mu=H_{0}$. For our
universe, another important correction may come from a boundary term
at the apparent horizon. Such a term may also change the scaling behavior
of the amplitude. 

If we assume that the Volume $V$ remains approximately the same after
we have added a matter amplitude $I_{m,eff}\propto\chi$, only the
Euler number can have changed if $\Lambda_{eff,s1}\neq\Lambda_{eff,s0}$, as there are no other quantities in eq. (\ref{eq:Lambdas12})
that could be responsible for a change of the saddle point. This suggests
that the physical interpretation of the $\chi$ dependent terms in
the effective matter action is simply to describe effects from black
or wormhole formation that one gets from energy fluctuations of the
matter. 

\section{Spacetime foam and the problem of time}

The models of Hawking and Barrow and Shaw both assume $\Lambda$ to
be a varying field as an axiom at the beginning. They do not give
an explanation why the cosmological constant should be considered
like this. Below, we will give arguments from quantum gravity which
suggest why the cosmological constant has to be found by a variational
principle and that it is related to boundary terms of instantons that
solve the problem of time. 

Hawking's spacetime foam model has a gravitational action $I=-\frac{\Lambda V}{8\pi}$
and thus, the classical gravitational background field contributes
\begin{equation}
S_{classical}\propto\ln\left(e^{-I}\right)\propto\frac{\Lambda V}{8\pi}
\end{equation}
to the gravitational entropy. In Hawking's work on blackhole entropy
from partition functions of gravity \cite{PartitionfunctionsHawking},
it is shown that the gravitational field produced by matter would
not yield any contribution to the entropy as long as no event horizon
is created. This implies that classical background spacetimes in Euclidean
quantum gravity can have entropy if and only if they have a boundary.
For Hawking's spacetime foam model, this means that $S_{classical}$
can only come from a boundary term. In the case of single blackholes,
one has a boundary at infinity and, since the Euclidean spacetime
can not describe the spacetime within the blackhole, one also gets
a boundary at the horzion. In the spacetime foam model, one assumes
that the spacetime is compact. So, there is no boundary at infinity.
However, we have found that for a given volume, the state of maximum
entropy has a background spacetime filled with $N=|c|V$ cavities. Therefore,
the boundaries that are responsible for the contribution $S_{classical}$
to the entropy must be associated with boundaries of the $N$ cavities.
Around the i-th cavity, we add the following Euclidean GHY boundary
term
\begin{equation}
I_{GHY,i}=\frac{1}{8\pi}\int_{\partial M_{i}}d^{3}x\sqrt{h}K_{i}^{ab}h_{ab}
\end{equation}
to the action $I_{\text{0}}=-\frac{1}{16\pi}\int d^{4}x\sqrt{g}R$
where $h_{ab}$ is the induced metric on the i-th boundary $\partial M_{i}$
and $K_{i}$ is the extrinsic curvature there. This proposal is actually
quite similar to a proposals made in \cite{Spacetimefoamccgerat}
and \cite{Xue}. The authors of \cite{Xue} show that this mechanism
would lead to inflationary behavior, and that it would even provide
an exit from inflation. Buth they do not observe that if the value
of the cosmological constant is given by the masses or areas of instantons,
then the cosmological constant should be found by a variational principle
that determines the most probable configuration for them. Below, we
will use the following approach: We have calculated the value of the
constant from the path integral in the section above, and now we will
compute the area of the instantons with this data, thus not needing
any assumptions about the black hole size like in \cite{Xue}. The
computation then will give us a physical explanation why a variational
principle has to be employed to compute the cosmological constant.

One should get with 
\begin{equation}
N=|\chi|=|c|V=|c|\int d^{4}x\sqrt{g}\label{eq:eevdevccdcc}
\end{equation}
 boundaries and the mean value 
\begin{equation}
\hat{I}_{GHY}=\frac{1}{N}\sum_{i=1}^{N}I_{GHY,i}\label{eq:dcccefejijoe}
\end{equation}
 an action
\begin{align}
I & =-\frac{1}{16\pi}\int d^{4}x\sqrt{g}R+\sum_{i=1}^{N}I_{GHY,i}\nonumber \\
 & =-\frac{1}{16\pi}\int d^{4}x\sqrt{g}R+N\hat{I}_{GHY}\nonumber \\
 & =-\frac{1}{16\pi}\int d^{4}x\sqrt{g}R+\hat{I}_{GHY}|c|\int d^{4}x\sqrt{g}\nonumber \\
 & =-\frac{1}{16\pi}\left(\int d^{4}x\sqrt{g}R-2\Lambda V\right),\label{eq:puroktrg}
\end{align}
where we have defined 
\begin{equation}
\Lambda\equiv8\pi |c|\hat{I}_{GHY}.\label{eq:xbjkqqqqq}
\end{equation}
Using Eqs. (\ref{eq:constantc}) and (\ref{eq:xbjkqqqqq}) one finds
that 
\begin{equation}
\hat{I}_{GHY}=\frac{106\pi}{3(73\Lambda+60\pi)}.\label{eq:wsgfdsffds}
\end{equation}
 If the boundaries of the cavities are similar to the boundaries of
event horizons from black holes in Euclidean quantum gravity, then
one would expect something like $\hat{I}_{GHY}=\frac{A}{4}$ where
$A$ is the area of each cavity. With $\Lambda\approx5.6\cdot10^{-122}$
one gets $A=4\hat{I}_{GHY}\approx2.37$ Planck areas.

In order to derive equations of motion, one writes the variation $\delta I$
as

\begin{eqnarray}
\delta I & = & \frac{-1}{16\pi}\delta\int d^{4}x\sqrt{g}R+\delta(|c|V\hat{I}_{GHY})\nonumber \\
 & = & \frac{-1}{16\pi}\delta\int d^{4}x\sqrt{g}R\nonumber \\
 &  & +|c|\hat{I}_{GHY}\int d^{4}x\delta\sqrt{g}+|c|V\delta\hat{I}_{GHY}\nonumber \\
 & = & \frac{-1}{16\pi}\delta\int d^{4}x\sqrt{g}R+\frac{\int d^{4}x\delta\sqrt{g}}{8\pi}\Lambda\nonumber \\
 &  & +\sum_{i=1}^{N}\int_{\partial M_{i}}d^{3}x\sqrt{h}\left(\delta K_{i}^{ab}h_{ab}\right)\nonumber \\
 & = & \frac{-1}{16\pi}\delta\int d^{4}x\sqrt{g}(R-2\Lambda),
\end{eqnarray}
where we have used the known fact that the sum of the boundary terms
$\int_{\partial M_{i}}d^{3}x\sqrt{h}\left(\delta K_{i}^{ab}h_{ab}\right)$
cancel another boundary term that one gets from the variation of $\frac{-1}{16\pi}\int d^{4}x\sqrt{g}g^{\mu\nu}\delta R_{\mu\nu}$.
We see that a contribution to the equations of motion which is similar
to the cosmological constant does arise because one has the the term
$\hat{I}_{GHY}$ for 
\begin{equation}
N=|c|V=|c|\int d^{4}x\sqrt{g}
\end{equation}
times in the action.

Eq. (\ref{eq:xbjkqqqqq}) also makes clear why the gravitational contribution
to the cosmological constant should be seen as a continuous field
whose value is to be found by a variational principle. The cosmological
constant $\Lambda=|c|8\pi\hat{I}_{GHY}$ is related to the mean of the
area (given by $\hat{I}_{GHY}$) and density (given by $|c|$) of the
cavities in the spacetime that are due to quantum fluctuations. In
general, the cavities can have any area and density per volume, which
would lead to different possible values of $\Lambda$. If we have
to sum the path integral over all possible metrics with different
boundaries or cavity densities, the system should then settle for
the configuration with maximum entropy. The equations which hold if
the entropy is at its maximum for a given volume are computed by $\frac{dN}{d\Lambda}=0$
and $\frac{dN}{d\chi}=0$.

The difference with ordinary black holes from Euclidean quantum gravity
appears to be mostly that the boundary is not due to the Wick rotation
since one should get a cosmological constant even when one is not
transforming the metric into an Euclidean one. It therefore can not
be an artifact of Euclideanization and one has to assume that the
boundaries are still there in the Lorentzian theory. In order to still
have these boundary terms after Wick rotation, we could assume that
spacetime is still simply connected and errect a wall around these
cavities that an observer can not cross or we could abandon the simply
connectedness. The negative Euler number that we computed from this model suggests that we should work with the latter option. The spacetime of Hawking's model allows the use of
the Hamiltonian of gravity, which implies that spacetime is globally
hyperbolic and can be written in the form of $\Omega=\mathbb{R}\times\Sigma$, where $\Sigma$ is a spacelike three manifold. Furthermore, in Hawking's
spacetime foam model, $\Sigma$ contains $N$ boundaries $\partial\Sigma$
and these boundaries should be there even in the Lorentzian spacetime
after Wick rotation. Now according to a definition given by Visser
in \cite{Visser}, if these conditions are met and if $\partial\Sigma\sim S^{2}$,
then $\Omega$ contains an intra universe wormhole. Therefore, the
cavities in Hawking's model is certainly compatible with wormholes, as the latter also imply that spacetime is not simply connected.

The action in Lorentzian spacetime is given by $S=\int dt\mathcal{L}$
where $\mathcal{L}$ is the Lagerangian density. After a wick rotation
to Lorentzian spacetime, the term $\Lambda V=|c|\hat{I}_{GHY}\int d^{4}x\sqrt{g}$
that was added to the Euclidean action gets a minus sign. If such
a term is subtracted from the Lorentzian action, then a term 
\begin{equation}
\mathcal{L_{\partial M}}=-|c|\hat{I}_{GHY}\int d^{3}x\sqrt{g}=-\frac{\Lambda V^{(3)}}{8\pi},
\end{equation}
where $V^{(3)}$ is the 3 volume, is subtracted from the Lageragian
density. The $\mathcal{L}_{\partial M}$ term is a boundary term because
$\Lambda=8\pi |c|\hat{I}_{GHY}$, where $\hat{I}_{GHY}$ is the average
of an area that comes from a boundary. For generalized coordinates
$q,p$, the Lagerangian density is given by $\mathcal{L}=\dot{q}^{i}p_{i}-H$
. Therefore, $\mathcal{L}_{\partial M}$ has to be added to the Hamiltonian.
As the Hamiltonian of general relativity without boundary term is
zero, the new Hamiltonian becomes: 
\begin{equation}
H=\frac{\Lambda V^{(3)}}{8\pi}\label{eq:Hamiltonian}
\end{equation}
or $tH=\frac{\Lambda V}{8\pi}$. This justifies Hawking's formula
for the canoncal partition sum in eq. (\ref{eq:wefewcwcewvccw}),
which should be $Z=tr\left(e^{-\tau H}\right)=tr\left(e^{-\frac{V\Lambda}{8\pi}}\right)$
with $\tau$ as a time coordinate in Euclidean space. A non zero Hamiltonian
leds, in accordance with the observations in \cite{deWitt,Feng} to
a solution of the problem of time in quantum gravity. 

By saying this, one should note that even if one can now write time
dependent operators like 
\begin{equation}
A(t)=e^{-itH}Ae^{itH},
\end{equation}
DeWitt argues in \cite{deWitt} that one only has time evolution if
the amplitude is not an Eigenstate of $H$ and one must construct
wave packets with different energy. Here we want to comment shortly
on parts of an article of Page and Wooters\cite{Page}. They begin
their letter on the problem of time by stating their unproven assumption
that there might exist a superselection rule for energy in quantum
gravity, similar to the superselection rules that exist for charge
in quantum electrodynamics. Such a rule would mean that there were
no superpositions with different energy and one would still be left
with a problem of time. To this, Hawking's spacetime foam model seems
to be a counter example. The states $|g_{n}\rangle$ are a complete
orthonormal base of energy Eigenstates according to Hawking. Since
one can define a trace over these states in eq. (\ref{eq:wefewcwcewvccw}),
it appears valid to build superpositions $|\Psi\rangle=\sum_{n}c_n|g_{n}\rangle$ with
them. Due to the equality of eqs. (\ref{eq:wefewcwcewvccw}) and (\ref{eq:Volumesummation})
some of the states $|g_{n}\rangle$ must be associated to different
4 volumes, i.e. they are eigenstates of the Wheeler DeWitt
equation for different three volumes and solve the Wheeler deWitt
equation at different times. Using $tH=\frac{\Lambda V}{8\pi}$ ,
one sees that eigenstates of the Wheeler DeWitt equation for different 3 volumes have different energy. The superposition $|\Psi\rangle$ then corresponds to a wave packet of eigenstates with different energy, as envisaged by DeWitt. 

The cosmological constant was computed by maximizing the entropy that
is given by the amplitude of Euclidean quantum gravity. Usually, there
is no time evolution in a system that has reached a state of thermodynamical
equilibrium and maximum entropy. However, we have computed this equilibrium
state only for a fixed given volume. If the system with a fixed volume
$V$ reaches a state of maximum gravitational entropy, it will have
a positive non vanishing cosmological constant of $\Lambda\approx H_{0}^{2}$.
On an infinitesimally small timescale, the latter implies an expansion
of the volume $V$ of the universe to $V+dV$. Using $\Lambda=\frac{-f}{\sqrt{V}}$,
the action can be written as $I=\frac{f}{8\pi}\sqrt{V}$. With $\Lambda>0$
and $V>0$, $f$ must be negative. Therefore, the action scales according
to $-I\propto\sqrt{V}$, and thus the entropy $S\propto-I\propto\sqrt{V}$
increases the larger the volume gets. 

At maximum entropy for a given volume, one has $|\chi|\propto V$. For
fixed Euler number, an increased $V$ would therefore bring the system
out of the maximum entropy condition and the universe must react by
creating new cavities. Once the system has created enough cavities
to reach a state of maximum entropy for the larger volume $V+dV$,
the universe will have a positive, non vanishing cosmological constant
as before, but that leads to further expansion. This expansion process
would only stop if one reaches $\Lambda=0$. According to the formula with the positive sign in eq. (\ref{eq:Lambdas12}), one can
have $\Lambda=0$ only in the limit of infinite volume. The timeless
equilibrium state of no expansion and maximum gravitational entropy
is therefore never reached exactly and the universe only approaches
it as it expands.

Because of $|\chi|\propto V$, the topology of the background spacetime
corresponding to the maximum entropy state should change during the
transition from a volume $V$ to$V+dV$. However, the spacetime foam
calculation does not describe the transition amplitudes that relate
the states of maximum entropy for different volumes to each other.
Unfortunately, there are difficulties in quantum gravity to describe
transition amplitudes. For example, the amplitude does not solve the
usual Wheeler deWitt equation anymore because one gets additional
terms in the Hamiltonian \cite{DeWitttransition, Feng}. Additionally, a result of DeWitt
\cite{Topology} states that topology changes of a classical background
spacetime would create singularities that would lead to infinite energy
production. This is a severe problem for the spacetime foam model
of Hawking, which shows that a maximum entropy state is given by a
different topology of the classical background spacetime for each
different volume and that the latter should expand in time.

With a spacetime made out of a gas of black or wormholes, one should
expect matter particles to scatter with the blackholes by getting
swallowed and replaced by outgoing radiation. It has been shown that in order to cross the
black hole horizon, the particle needs to be accelerated to high
energy, as otherwise it would be in an inertial system where it would
see a Poynting-Robertson effect dragging it away from the blackhole
\cite{horizoncross}. This seems especially to be the case for micro blackholes. The lifetime $\tau$ of an isolated blackhole
is approximately
\begin{equation}
\tau\propto M^{3},\label{eq:Lifetime}
\end{equation}
see \cite{DeWittcurved}. With $M=2R$ and $R$ being the Schwarzschild
radius, $\tau$ is therefore very short for a blackhole if $R$ is
approximately around one Planck length. Setting $\tau=\Delta t$ into
\begin{equation}
\Delta E\Delta t\geq\hbar/2
\end{equation}
shows that one would need to detect processes of very high energy
if one would be able to observe the process of a point particle flying
into a Planck sized blackhole and the particle getting replaced by
outgoing radiation. 

The usual Feynman rules of quantum field theory appear only to include
the scattering of particles with themselves. They do not to include
rules for scattering processes of a particle with the event horizon
of a blackhole. However, with a result of 't Hooft,
the amplitude of string theory may be interpreted as the scattering
matrix of exactly such a process. In 't Hooft's model, a particle's gravitational field is described by a Schwarzschild metric. Upon falling into a blackhole, the particle accellerates to relativistic speed and generates a shockwave \cite{Thooft2}. This modifies the motion of the outgoing particles and the resulting scattering amplitude is the usual amplitude over the Polyakov action from string theory (in Wick rotated form) \cite{Thooft}.

A particle that flies through a
spacetime foam made out of a gas of worm- or black holes has to encounter
many of these scattering processes. Moreover, one also has to expect
blackhole production whenever particles of very high energy collide.
Therefore, one has to take the process of a particle that scatters
with blackholes into account if one wants to write any amplitude which
should be valid at very high energies. This argument also applies
for gravity itself. Hence, one can expect that one has to use string
theoretic corrections for high energy graviton scattering. As a result,
a more precise description of the dynamics of spacetime foam might
be given in the future within the framework of string theory. For
string theory amplitudes, it was possible to demonstrate that at least
in some of its extra dimensions one can have topological transitions\cite{topologychange}.
Perhaps this opens a way to cure the remaining problems of Hawking's
spacetime foam model.

\section{How spacetime foam may influence the equations of motion of matter}

\subsection{Non relativistic case}

Quantum fluctuations of gravity might have experimentally observable
consequences if they describe a spacetime foam. After Hawking wrote
his first article on this idea, Hawking, Page, Pope and Warner immediately
published works where they investigated how quantum mechanical particles
might change their trajectory if they were put close to a virtual
black hole from spacetime foam \cite{Bubble1,Bubble2,Bubble3}. They
computed the S matrix of quantum mechanical particles e.g. in a $CP^{2}$
and a $S^{2}\times S^{2}$ topology and found large corrections for
scalar particles that would forbid the existence of the Higgs boson,
which has now been found at the LHC. 

Above, we noted that boundary terms have to be associated with the
gravitational instantons of spacetime foam. The effects of boundary
terms on the particle behavior were not investigated by Hawking and
coworkers. One effect of them is that the particle can never be observed
coming close to a region of very large curvature. This may cure some
of the problems with the amplitudes from \cite{Bubble1,Bubble2,Bubble3}.
Unfortunately, the boundary terms create a severe additional problem
as they should lead to Hawking radiation. Far away from a single black
hole of mass $M=2R$, where $R$ is the Schwarzschildradius, an observer
should see a photon gas of a temperature 
\begin{equation}
T_{H}=\frac{hc^{3}}{8\pi GMk_{b}}\label{eq:hawkingtemperature}
\end{equation}
with an infinite number of photons. A spacetime where a high number
of planck sized black or wormholes with event horizons are spontaneously
produced from quantum fluctuations would therefore create a photon
gas of high temperature and particle number. In such a photon gas,
matter particles should behave differently as in vacuum due to repeated
compton scattering with the photons. This is what we will investigate
below. 

For now, we will work on a low energy scale and a large time intervall.
Then, as we mentioned above, one can not observe the particle coming
close to the black hole and get swallowed, since after a very short
evaporation time, any information swallowed by the black hole should
be restored, and the blackhole should be gone. Spacetime should be
homogeneous and therefore, we assume the blackholes are distributed
randomly over the entire spacetime. From far away, the gravitational
field of a gas of randomly distributed black holes is not different
from the gravitational field of a gas of stars and one can use Newtonian
mechanics at large distances in both cases. By a calculation of Chandrasekhar
\cite{chandra}, a classical particle of mass $m$ that moves with
velocity $\dot{\mathbf{\sigma}}(t)$ through such a gravitational
field would be subjected to a fluctuating force that leads, on long
time scales, to a gravitational drag according to Stokes law:
\begin{equation}
\mathbf{F}_{v}=m\ddot{\mathbf{\mathbf{\mathbf{\sigma}}}}(t)=-\gamma\mathbf{\dot{\mathbf{\mathbf{\sigma}}}}(t)\label{eq:frictionalforce}
\end{equation}
The friction coefficient $\gamma$ is of the order of the reciprocal
time $\tau$ of relaxation, which is the time how long it takes for
cumulative effects to have an influence over the 2 body interaction.
If one assumes that the cavities of spacetime foam have a Schwarzschildradius
$R=2M$ of approximately Planck length and one could use the small
quantity $\tau\propto M^{3}$ as relaxation time, then one would have
to expect a very large friction$\gamma\propto1/\tau$. It was noted by Liberati and Maccione in \cite{viscosity}
that the naive assumption of a vacuum with a large friction
coefficient would have severe consequences for matter particles that
should be experimentally measurable. In the following we will argue
why the effects described in \cite{viscosity} are not observed.

In his article \cite{chandra}, Chandrasekhar computes an explicit
formula for the friction coefficient $\gamma$ based on the assumptions
that a particle would be in purely gravitational interaction with
a random field of stars and could orbit a single star for a relaxation
time of $\tau$. For the case of a test particle orbiting a supermassive
blackhole, Chandrasekhar's calculation was adapted in \cite{Friction2}
and the result is a velocity independent friction coefficient and
a stokes law in the form of (\ref{eq:frictionalforce}). Unfortunately,
this computation for $\gamma$ is not entirely suitable for the case
of micro blackholes as it does not take effects of their radiation
into account. Therefore, we have to adapt our model for this situation. 

Hawking radiation consists of photons of all frequencies $\nu$. On
classical matter particles, $N$ photons induce a force of magnitude
\begin{equation}
|\mathbf{F}|=\nu\frac{dN(t)}{dt}.\label{eq:pressure}
\end{equation}
As the photon gas is created by spontaneously emerging and exploding
black holes, one should have density fluctuations of this gas. For
a blackhole with Schwarzschildradius $R_{0}$, the number of photons
that pass a far away spherical surface of radius $r>R_{0}$ is given
by 
\begin{equation}
<N>=\frac{1}{2\pi}\frac{\Gamma(\omega)}{e^{\omega/k_{b}T}-1}\label{eq:Particlenumber}
\end{equation}
 where $\omega$ is the angular frequency of the radiation per unit
time and $\Gamma(\omega)$ is an absorbtive coefficient. If we use
the temperature of eq. (\ref{eq:hawkingtemperature}) in eq. (\ref{eq:Particlenumber}),
then with 
\begin{equation}
\frac{dM}{dt}=\propto-\frac{1}{M^{2}}\label{eq:decaylaw}
\end{equation}
from \cite{DeWittcurved} one can try to get an estimate of (\ref{eq:pressure})
for a particle that orbits an isolated blackhole far away. From this,
one would have to expect that the photon gas from black holes that
emerge and explode at random places in spacetime creates a random
force $\mathbf{F}^{r}(t)$ that acts at every point in space on a
traversing particle. We assume that the black holes of spacetime foam
are equally distributed in space. This means that if at a certain
time, Hawking radiation of a blackhole induces a force $\nu\frac{dN(t)}{dt}$
in $\vec{x}$ direction on a particle, another blackhole could later
induce a force $-\nu\frac{dN(t)}{dt}$ on the same particle in the
same direction. For all directions, this would imply that the average
of $\mathbf{F}^{r}$ vanishes, or
\begin{equation}
<\mathbf{F}^{r}>=0.
\end{equation}
Furthermore, we assume for now that the appearance and explosion of
several of these blackholes should not be correlated events in time.
Hence, $\mathbf{F}^{r}(t)$ should be uncorrelated with $\mathbf{F}^{r}(t-1)$.
All this implies by a standard argument invoking the central limit
theorem, see \cite{Lennart}, that $\mathbf{F}^{r}(t)$ should be
a Wiener process with a Gaussian distribution. With the random force
term and the friction term acting on the particle, we get Langevin's
famous equation:

\begin{eqnarray}
m\ddot{\mathbf{\sigma}}(t)+\gamma\mathbf{\dot{\mathbf{\sigma}}}(t) & = & \mathbf{F}^{r}(t)\label{eq:brownian1-1}
\end{eqnarray}
and since $\mathbf{F}^{r}(t)$ is Gaussian, we can use all the results
from the classical theory of Brownian motion.

If we want to compute e.g. a probability density of the particle to
be at a certain time and position in a gas of black holes, it may
help to use the ergodic hypothesis and consider instead of a single
trajectory a statistical ensemble of $j$ trajectories with velocity
$\dot{\mathbf{\sigma}}_{j}$. By considering the motion of the system
without $\mathbf{F}_{j}^{r}(t)$ one gets $\gamma\propto\frac{m}{\tau}$
with $\tau$ as relaxation time. One may additionally note that the
experiment may consist of additional forces $\mathbf{F}^{ext}$ as
well. One then gets
\begin{eqnarray}
m\ddot{\mathbf{\sigma}}_{j}(t)+\frac{m}{\tau}\mathbf{\dot{\mathbf{\sigma}}}_{j}(t) & = & \mathbf{F}_{j}^{r}(t)+\mathbf{F}^{ext}.\label{eq:brownian1}
\end{eqnarray}
From this, one may compute averages over $j$ that describe the single
system observed over some time. 

It is known that gravitational potentials of a star can reduce the
entropy of a surrounding gas of particles by compressing the gas and
slowing it down (the compression in the gravitational field also heates
the gas but this effect is smaller than the entropy reduction from
the reduction of the particle motion, see \cite{Baez}. The matter
that is slowed down then gets on a trajectory where it falls into
the star whose entropy increases). In our case, the dynamical friction
induced by the gravitational fields collectively reduces the entropy
of a particle ensemble. Thereby, eq. (\ref{eq:brownian1}) contains
all the ingredients of the black hole information paradox: A term
that reduces the entropy of a particle system, and a random noise
term that is purely thermal and does not suffice to restore the original
state of the particles, as it does not contain any information about
it.

The black hole entropy generated by Hawking radiation is much larger
than the entropy of the surrounding matter. A proposal by DeWitt \cite{Globalap}
is therefore that the infalling particles interact with the outgoing
Hawking radiation and modify its Bogoliubov modes. This interaction
may restore the state of the surrounding matter. If the information
is preserved upon black hole decay, one has to expect that the friction
term in eq. (\ref{eq:brownian1}), which reduces the entropy of the
particles, gets reversed into $-\gamma\mathbf{\dot{\mathbf{\mathbf{\sigma}}}}_{j}(t)$
at a later stage of the process by the radiation. A corrected radiation
term 
\begin{equation}
\tilde{\mathbf{F}}_{j}^{r}(t)=2\frac{m}{\tau}\mathbf{\mathbf{\dot{\mathbf{\sigma}}}}_{j}(t)+\mathbf{F}_{j}^{r}(t)\label{eq:brownian2-1}
\end{equation}
should then restore the original state of the matter. One gets the
following equation: 
\begin{equation}
m\ddot{\mathbf{\sigma}}_{j}(t)-\frac{m}{\tau}\mathbf{\mathbf{\dot{\mathbf{\sigma}}}}_{j}(t)=\mathbf{F}_{j}^{r}(t)+\mathbf{F}^{ext}.\label{eq:brownian2}
\end{equation}

Although they did not connect eqs. (\ref{eq:brownian1}) and (\ref{eq:brownian2})
to the behavior of a particle ensemble that is put into a gas of black
hole like objects, these equations were first proposed by Fritsche
and Haugk in \cite{Fritsche} as a starting point to derive Schroedinger's
equation. 

In their calculation, they separate the $\dot{\mathbf{\sigma}}_{j}$
into $\dot{\sigma}_{j}(t)=\dot{\mathbf{\sigma}}_{jr}(t)+\dot{\mathbf{\sigma}}_{jc}(t)$,
where $\dot{\mathbf{\sigma}}_{jc}(t)$ is the convective velocity
that would occur if $\mathbf{F}_{j}^{r}$ would be absent, and $\dot{\mathbf{\sigma}}_{jr}(t)$
is caused by the random force. Fritsche and Haugk then computed computed
ensemble averages of equations (\ref{eq:brownian1}) and (\ref{eq:brownian2})
over $j$ and get 
\begin{equation}
\partial_{t}(\mathbf{v}-\mathbf{u})+(\mathbf{v}+\mathbf{u})\nabla(\mathbf{v}-\mathbf{u})-\nu\Delta(\mathbf{v}-\mathbf{u})=\frac{1}{m}\mathbf{F}^{ext}\label{eq:fsfdsdv}
\end{equation}
 and 
\begin{equation}
\partial_{t}(\mathbf{v}+\mathbf{u})+(\mathbf{v}-\mathbf{u})\nabla(\mathbf{v}+\mathbf{u})+\nu\Delta(\mathbf{v}+\mathbf{u})=\frac{1}{m}\mathbf{F}^{ext}.\label{eq:ycycyc}
\end{equation}
In the equations above, $T$ is the effective temperature of the heath
bath, $\mathbf{v}=\mathbf{v}_{c}+\mathbf{u}$ the ensemble average
of $\dot{\sigma}_{j}$ over j and $\mathbf{v}_{c}$ is the ensemble
average of $\dot{\sigma}_{jc}$. Furthermore, $\mathbf{u}=-\nu\frac{1}{\rho}\nabla\rho$
where $\rho$ is the probability density of the particle and $\nu=\frac{k_{b}T\tau}{m}$
is a diffusion coefficient which one may set to $\nu\equiv\frac{\hbar}{2m}.$

Computing then the average of eqs. (\ref{eq:fsfdsdv}) and (\ref{eq:ycycyc}),
one arrives at
\begin{equation}
\frac{d}{dt}\mathbf{v-\mathbf{(u}}\nabla)\mathbf{u}+\nu\Delta\mathbf{u}=\frac{1}{m}\mathbf{F}_{ext}.\label{eq:ycycyc-1}
\end{equation}
Setting $\hbar=0$ results in $\nu=0$ and Newton's second law $\mathbf{F}=m\dot{\mathbf{v}}$.
On the other hand, with $\hbar\neq0$, one can derive the one particle
Schroedinger equation 
\begin{equation}
i\hbar\partial_{t}\psi=\left(\frac{-\hbar^{2}\nabla^{2}}{2m}+V_{ext}\right)\psi
\end{equation}
from eq. (\ref{eq:ycycyc-1}).

If we could use eq. (\ref{eq:hawkingtemperature}) in 
\begin{equation}
\hbar=2k_{b}T\tau,\label{eq:hbar}
\end{equation}
it would imply that 
\begin{eqnarray}
\tau & = & \frac{\hbar}{2k_{b}T}=\frac{4\pi GM}{c^{3}}.\label{eq:relaxationtime}
\end{eqnarray}
For a Schwarzschild black hole 
\begin{equation}
M=\frac{R_{0}c^{2}}{2G},
\end{equation}
and with a Schwarzschildradius of Planck length $R_{0}=l_{p}$, this
would mean a relaxation time of $\tau=\frac{2\pi l_{p}}{c}=3.35\cdot10^{-43}s$,
which is shortly above Planck time of $5.3\cdot10^{-44}s$. We note
that with these assumptions, neither $\tau$ nor $\hbar$ do depend
on the gravitational constant. Hence, no matter if we live in a universe
where $G$ is small or not, if the heath bath is produced by Hawking
radiation of a gas of black holes of Planck size, one would always
get the same relaxation time and Planck's constant. However, one should
note that eq. (\ref{eq:hawkingtemperature}), the temperature one
sees for a single blackhole at infinity can, strictly speaking, probably
not be used to compute the temperature in eq. (\ref{eq:hbar}). The
temperature in eq. (\ref{eq:hbar}) contains the average temperature
that a particle sees in a gas of randomly occuring blackholes that
are coupled to a heath bath of their own radiation. This effective
temperature is defined in the Gaussian probability distribution of
$\mathbf{F}_{j}^{r}(t)$ which is given by 
\[
P(\mathbf{F}_{j}^{r})=\frac{1}{\sqrt{\pi}\frac{1}{\tau_{coll}}\sqrt{mk_{b}T}}e^{-\left(\mathbf{F}_{j}^{r}/\left(\frac{1}{\tau_{coll}}\sqrt{mk_{b}T}\right)\right)^{2}}
\]
where $\tau_{coll}$ is a mean time of momentum transfer from the
encounters with thermal photons. 

Certainly, the effective average temperature that comes from random
encounters with the radiation of many blackholes does not correspond
exactly to the temperature that an observer sees for a single isolated
blackhole at infinity. But perhaps one can use the latter temperature
as an approximation for particles which are never observed to come
close to the black holes of spacetime foam. To the author, the reversible
diffusion process outlined above appears to be a possible explanation
for why we do not observe any of the effects mentioned in \cite{viscosity}
when we are dealing with a gas of Planck sized black holes that would,
according to Chandrasekhar's calculations, give rise to a highly viscuous
medium with viscosity coefficient $\gamma\propto\frac{1}{\tau}$ for
all particles immersed in it.

This is certainly only a first idea to solve the problems that spacetime
foam poses for the behavior of matter particles. It is, for example
still unclear, how one should derive the correlations observed in
entangled states from such a model. How to extend these ideas relativistically
is also unclear at the moment.

\end{document}